\documentclass[fleqn,12pt,twoside]{article}
\usepackage{espcrc1}

\usepackage{graphicx}
\usepackage{epsfig}    
\usepackage[figuresright]{rotating}

\newcommand{\AmS}{{\protect\the\textfont2
  A\kern-.1667em\lower.5ex\hbox{M}\kern-.125emS}}
  
\def\funp{{I\!\!P}}
\def\xp{x_{{I\!\!P}}}

\newcommand{\be}{\begin{equation}}
\newcommand{\ee}{\end{equation}}
\newcommand{\beeq}{\begin{eqnarray}}
\newcommand{\eeeq}{\end{eqnarray}}

\title{Theoretical review of diffractive phenomena}

\author{K. Golec-Biernat
\address{H. Niewodnicza\'nski Institute of Nuclear Physics,
Polish Academy of Sciences \\ 31-342 Cracow, Poland}
}
       
\begin{document}

\maketitle

\begin{abstract}
We review QCD based descriptions of diffractive deep inelastic scattering
emphasising the role of models with  parton saturation.
These models provide natural explanation of such experimentally observed facts as the constant
ratio of $\sigma^{diff}/\sigma^{tot}$ as a function of the Bjorken variable $x$, and Regge
factorization of diffractive parton distributions. The Ingelman-Schlein
model and the soft color interaction model are also presented.

\end{abstract}

\section{Introduction}

One of the most important experimental results from the DESY $ep$ collider HERA
is the observation of a significant fraction (around $10\%$) of diffractive events
in deep inelastic scattering (DIS) with
large rapidity gap between  the scattered proton, which remains intact,
and the rest of the final system \cite{H197,ZEUS99,ZEUS04}.
In the standard, QCD description
of DIS  such events are not expected in such an abundance since large
gaps are exponentially suppressed due to color strings formed between
the proton remnant and scattered  partons. For diffractive events, however,
a color neutral cluster of partons fragments independently of the scattered proton.
The ratio of diffractive to all DIS events depends weakly on the  Bjorken variable $x$
and photon virtuality $Q^2$. Thus, DIS diffraction is a leading twist effect with
logarithmic scaling violation in $Q^2$. The theoretical description of diffractive events is a
real challenge since it must combine perturbative QCD effect of hard scattering with
nonperturbative phenomenon of rapidity gap  formation. It would be also desirable to
apply this description to analogous diffractive phenomena in hadronic collisions
with hard jets separated in rapidity from  (one or two) unshattered  hadrons. Actually,
hard diffraction was observed for the first time 
in $p\bar{p}$ scattering by UA8 collaboration \cite{UA8}.

In this presentation we concentrate on the discussion of theory of hard diffraction, when
there exists a hard scale, the photon virtuality $Q^2$ or  jet transverse momentum,  which
allows to apply perturbative QCD. Soft diffraction, when such a scale is missing, is outside the scope of our review. The reason being no significant progress in the development of new theoretical ideas concerning soft diffraction since the seventies,
in addition to the existing ones based on the Regge pole phenomenology. This phenomenology, however, turns out to be quite useful in the description of a soft part of hard diffraction, responsible for the rapidity gap formation.

\section{Diffractive parton distributions}

\begin{figure}[t]
  \vspace*{-0.5cm}
     \centerline{
         \epsfig{figure=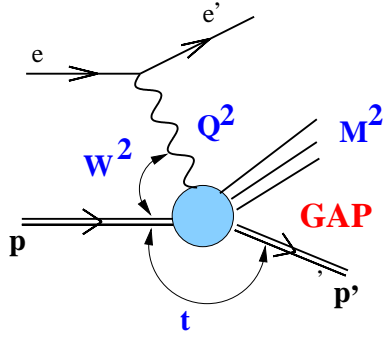,width=5cm}
           }
\vspace*{-0.5cm}
\caption{\it Kinematic invariants in DIS diffraction in  electron--proton collision.
\label{fig:d1}}
\end{figure}

Let us start with a brief description of kinematical variables in DIS diffraction,
shown in Fig.~\ref{fig:d1}. In addition to the photon virtuality $Q^2$ and
total energy of
the $\gamma^*p$ system $W$ , there are two additional invariant variables related to the diffractive
nature of the process: invariant mass of the diffractive system $M^2$ and the squared momentum
transfer $t$.  The following dimensionless fractions are built 
out  of these variables:
\be
\label{xp}
\xp \;=\; \frac{Q^2+M^2-t}{Q^2+W^2}\,,
\ee
which is a fraction of the incident proton momentum transferred into the diffractive system, and
\be
\label{beta}
\beta \;=\; \frac{Q^2}{Q^2+M^2-t}\,,
\ee
being an analogue of the Bjorken variable $x$ for the diffractive system. Experimentally
$|t|\ll Q^2,M^2$, thus $t$ can be neglected in the above formulas.
Finally, the Bjorken variable
\be
\label{x}
x\;=\;\frac{Q^2}{Q^2+W^2}\;=\;\beta\,\xp\,.
\ee
After averaging over the azimuthal angle of the scattered proton, the diffractive cross
section is characterized by two dimensionful {\it diffractive structure functions} $F_{2,L}^{D(4)}$
\be
\label{eq:difsf}
\frac{d^4\sigma^D}{dx\, dQ^2\, d\xp\, dt}
\,=\,
\frac{2\pi \alpha^2_{em}}{x\, Q^4}
\left\{
\left[1+(1-y)^2\right] F_{2}^{D(4)}
-\,{y^2}\,F_{L}^{D(4)}
\right\}\,,
\ee
in a full analogy to inclusive DIS. They depend on four variables: 
$(x, Q^2; \xp, t)$. After the integration over $t$ (if $t$ is not measured), 
the dimensionless structure functions are obtained. Due to the kinematical factor
in (\ref{eq:difsf}), we neglect the
longitudinal structure function $F_{L}^{D(4)}$ in the following.

The leading twist description of diffractive DIS is realized using {\it diffractive
parton distributions} (DPD) $q^D_i$, where $i$ enumerates quark flavour, in terms of which
\be
\label{eq:2}
F^{D(4)}_{2}=\sum_{i=1}^{N_f}e_i^2\,\beta\,\left\{q^D_i(\xp,t;\beta,Q^2)+
\overline{q}^D_i(\xp,t;\beta,Q^2)\right\}\,,
\ee
in the leading logarithmic approximation. In addition to the quark distributions,
 the gluon DPD $g(\xp,t;\beta,Q^2)$ is also defined.
Eq.~(\ref{eq:2}) is an example of the collinear factorization formula
proven for DIS diffraction in \cite{COL98}.
In the infinite momentum frame, the DPD have an interpretation
of conditional probabilities to find a parton in the proton
with the momentum fraction $x=\beta \xp$ under the condition that
the incoming proton stays intact and loses
the fraction $\xp$ of its  momentum.
A systematic approach to diffractive parton distributions, based
on quark and gluon operators, is given in \cite{BERSOP96,HAUT98}.

The $Q^2$-dependence of DPD is governed by the Altarelli-Parisi (DGLAP)
evolution equations. In order to find this
dependence, initial conditions at some starting scale have to be specified,
e.g. from
fits to diffractive DIS data in full analogy to the inclusive case
\cite{H197,ZEUS99,ZEUS04}.
In the evolution equations only $(\beta, Q^2)$ are relevant variables while
$(\xp,t)$ play the role of external parameters. Thus, a modelling of the latter dependence for
DPD is necessary. This is done using physical ideas about the nature of 
interactions leading to DIS diffraction.

\begin{figure}[t]
  \vspace*{-0.5cm}
     \centerline{
         \epsfig{figure=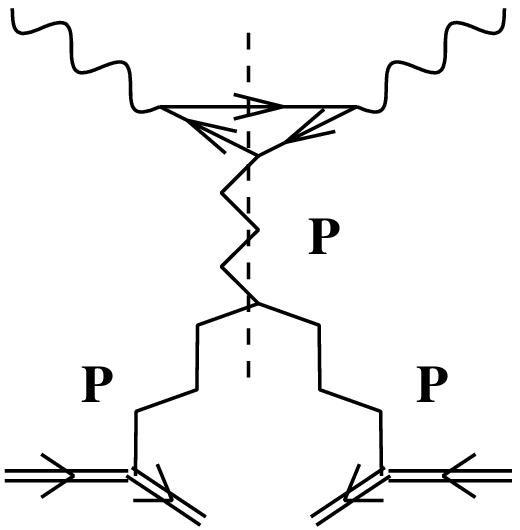,width=4cm}
         \hskip 2cm
         \epsfig{figure=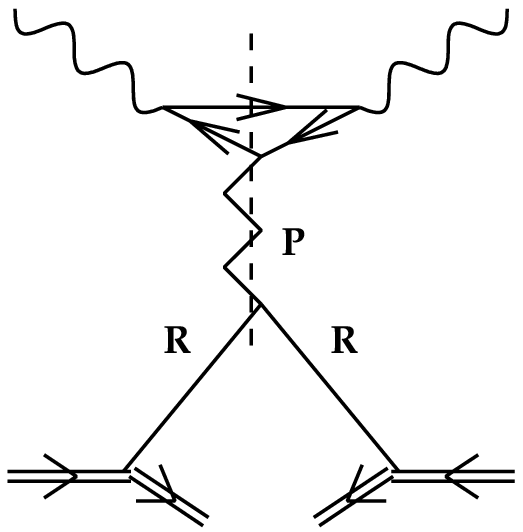,width=4cm}
           }
\vspace*{-0.5cm}
\caption{\it The pomeron and reggeon contributions to diffractive structure function.
\label{fig:dreg}}
\end{figure}

Traditionally, diffraction is related to  the exchange of a pomeron. This is 
the dominant at high energy  vacuum quantum number exchange,
described by a Regge pole with  a linear trajectory
$\alpha_\funp(t)=\alpha_\funp(0)+\alpha^\prime\, t$ and the intercept
$\alpha_\funp(0) \ge 1$. In  the Ingelman--Schlein
model \cite{IS} of hard diffraction the  pomeron exchanged between the proton and  diffractive system is supplemented by hard QCD stucture with partons.
In this case, the DPD  factorize into a pomeron flux
\be
\label{eq:pomflux}
f(\xp,t)\,=\,
\frac{B^2(t)}{8\pi^2}\;\xp^{1-2\alpha_\funp(t)}\,,
\ee
and pomeron parton distributions $q^\funp_i(\beta,Q^2)$:
\be
\label{eq:Reggefact}
q^D_i(\xp,t;\beta,Q^2)=f(\xp,t)\,q^\funp_i(\beta,Q^2)\,.
\ee
In the above $B(t)$ is the Dirac electromagnetic form factor of the proton \cite{DL84}, and
$\beta$ is a fraction of the pomeron momentum carried by a struck quark.
Since the pomeron carries the vacuum quantum numbers, the pomeron quark and antiquark distributions are equal: $q^\funp_i=\overline{q}^\funp_i$.
The inspired by the Regge theory factorization (\ref{eq:Reggefact})
is called {\it Regge factorization}. To good accuracy,
this type of factorization was found in the diffractive date at HERA \cite{H197,ZEUS99}.

The QCD analysis of the early diffractive data form HERA, using the Ingelman-Schlein  model,
was done in \cite{GK} with the
soft pomeron trajectory $\alpha_\funp(t)=1.1+0.25\cdot t$ and
parameters of the pomeron parton distributions determined from analyses of
soft hadronic reactions.
More recent  analyses of inclusive DIS diffraction
\cite{H197,ZEUS99,ZEUS04} assume  Regge form of DPD (\ref{eq:Reggefact})
determined by the DGLAP based fits. In particular, the effective slope $\alpha_\funp(0)=1.16$
was found as a result of a fit in the recent analysis \cite{ZEUS04}.
In all cases, the fits give large gluon DPD with the relative contribution
to the pomeron momentum around $80-90\%$ \cite{H197}.

The collinear factorization fails in hadron--hadron hard diffractive scattering
due  to initial  state soft interactions \cite{CFS93,REVWU}. Thus, unlike
inclusive scattering,  the diffractive parton distributions are no universal
quantities. They can be used, however,  for different diffractive processes in DIS
scattering, e.g. diffractive dijet production,  for which the collinear factorization
theorem holds.

\section{Subleading reggeons}

\begin{figure}[t]
 \vspace*{-1.0cm}
     \centerline{
         \epsfig{figure=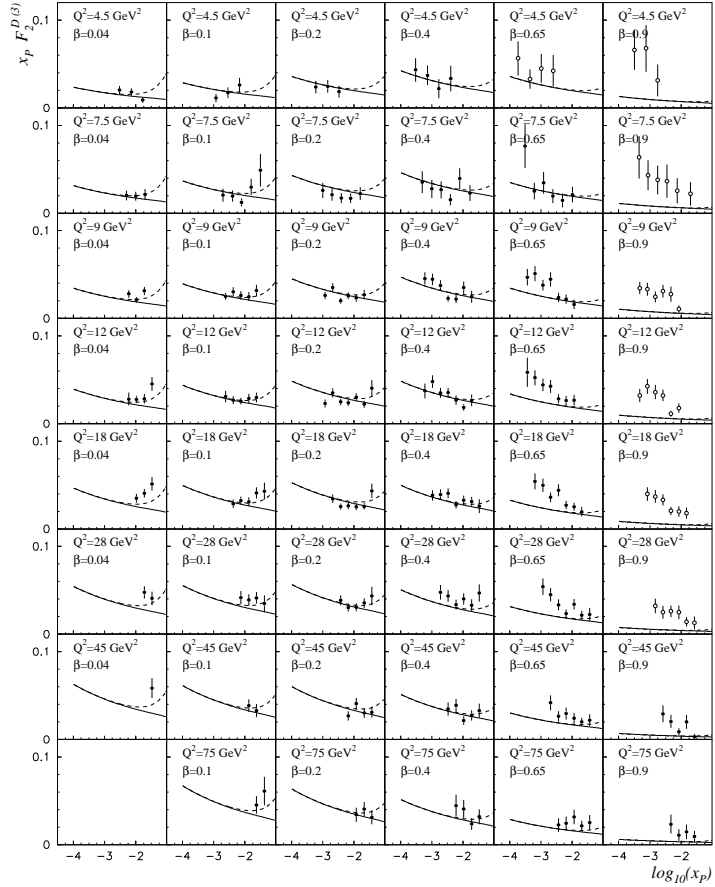,width=10cm}
           }
\vspace*{-0.5cm}
\caption{\it The pomeron (solid) and reggeon (dashed) contributions
to diffractive structure function \cite{GBK2}. The data are from \cite{H197}.
\label{fig:ddreg}}
\end{figure}

The exchange of subleading reggeons can account for the Regge factorization breaking
of diffractive structure function
for large values of $\xp>0.01$. Strictly speaking, we cannot call such processes diffractive
since  diffraction is usually associated with the leading pomeron exchange.
However, for simplicity we use the same terminology for the non-pomeron exchanges,
including the isospin changing process with neutron instead of the proton  in the final state.
The reggeon contribution is shown in Fig.~2, which illustrates the following extension of the
Ingelman-Schlein model \cite{GBK2}
\be
\label{eq:f2d4r}
F_2^{D(4)}(x,Q^2,\xp,t)\;=\;f_\funp(\xp,t)\,F_2^{\funp}(\beta,Q^2)
\,+\,
\sum_R f_R(\xp,t)\,F_2^{R}(\beta,Q^2)\,,
\ee
where the non-pomeron terms describe reggeon exchanges, isoscalar $(f_2,\omega)$ and isovector $(a_2,\rho)$, with the  trajectory $\alpha_R(t)=0.5475+1  \cdot t$
in the reggeon fluxes $f_R(\xp,t)$. $F_2^{R}$ is a reggeon structure function determined in
\cite{GBK2}.
With such a structure function the Regge factorization is obviously broken for large
$\xp$, which is shown in Fig.~3 by dashed lines.

\section{Parton saturation and diffraction}

\begin{figure}[t]
  \vspace*{-1.5cm}
     \centerline{
         \epsfig{figure=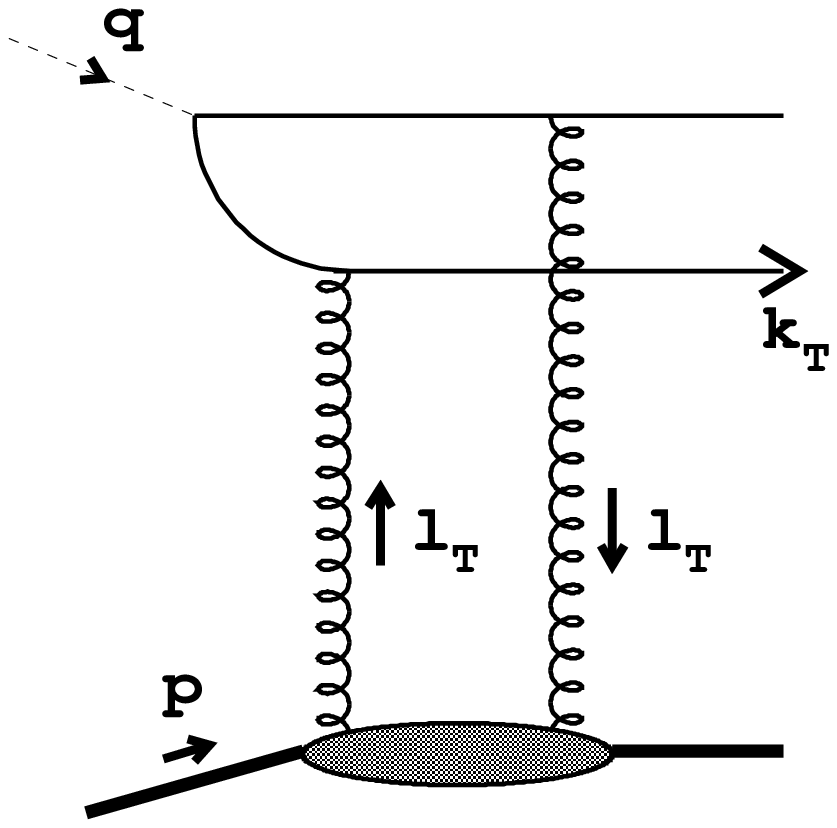,width=5.5cm}
         \epsfig{figure=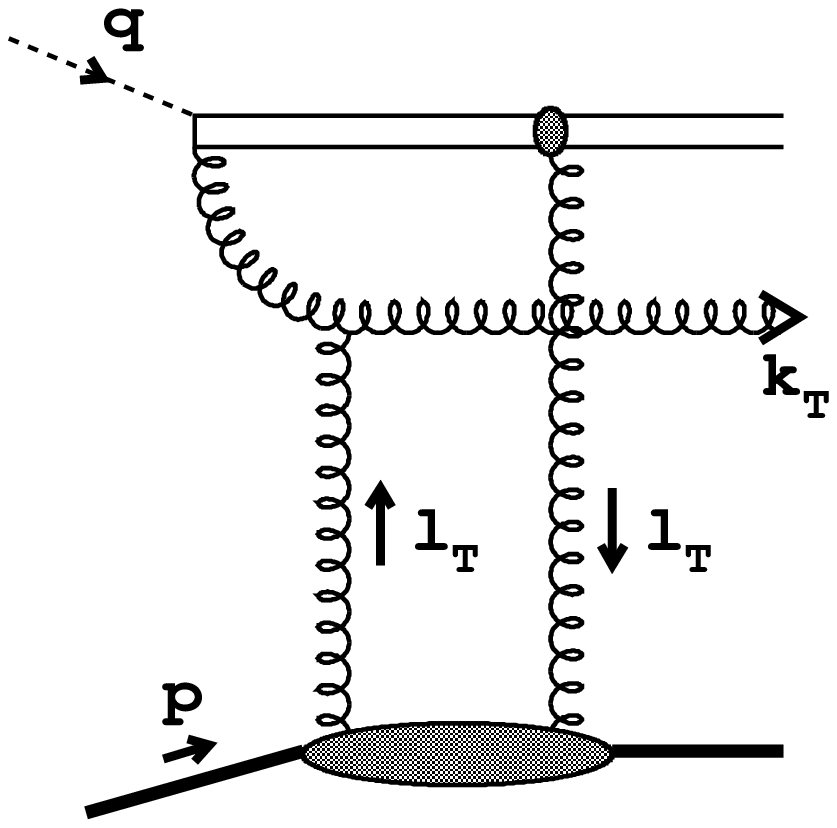,width=5.5cm}
           }
\vspace*{-1cm}
\caption{\it The $q\bar{q}$ and $q\bar{q}g$ components of the diffractive system.
\label{fig:diag1}}
\end{figure}

The leading twist DPD lead to
good description of data.  However, the basic expe\-ri\-men\-tal fact that
$\sigma^{diff}/\sigma^{tot}\simeq const$ as a function of energy $W$
is not understood in this approach.
The understanding is provided in a different theoretical framework of DIS diffraction
in which the virtual photon splits into a quark-antiquark pair 
which subsequently scatters off the target proton through a further quantum fluctuation.
This picture is valid in the frame in which the $q\bar{q}$ pair (dipole) carries most of
the available rapidity $Y\sim \ln(1/x)$ of the system, and the light-cone photon momentum $q^+>0$.
The gluon radiation from the parent dipole can be interpreted in the large $N_c$ limit
as a collection of dipoles of different transverse sizes which interact with the proton. If the proton stays intact, the diffractive events with large rapidity gap are formed. In such a case, the diffractive system is given by the color dipoles and the pomeron
can be modelled by color singlet gluon exchange between
the dipoles and the proton.

In the simplest case when only the parent $q\bar{q}$ dipole form a diffractive system,
see Fig.~\ref{fig:diag1}, the diffractive cross section at $t=0$ reads \cite{NN}
\be
\label{eq:5}
\frac{d\,\sigma^{diff}}{dt}_{\mid\, t=0}
\,=\,
\frac{1}{16\,\pi}\,
\int d^2 r\, dz\,
|\Psi^\gamma(r,z,Q^2)|^2\ \hat\sigma^2(x,r),
\ee
where   $\Psi^\gamma$ is the well known light-cone wave function of the virtual photon,
$r$ is the dipole transverse size and $z$ is a fraction of the photon
momentum $q^+$ carried by the quark. The {\it dipole cross section} $\hat\sigma(x,r)$ in this formula
describes the pomeron interaction, which in the QCD approach is modelled by the exchange of gluons.
The simplest, two gluon exchange does not depend on energy and has to be rejected. Since
the DIS difraction is a typical high energy (small $x$) phenomenon, it is tempting to apply
the BFKL pomeron \cite{BFKL} with two reggeized, interacting gluons. However, the resulting energy dependence is too strong in this case. Thus, more complicated gluon exchanges are necessary.

\begin{figure}[t]
  \vspace*{-1.0cm}
     \centerline{
         \epsfig{figure=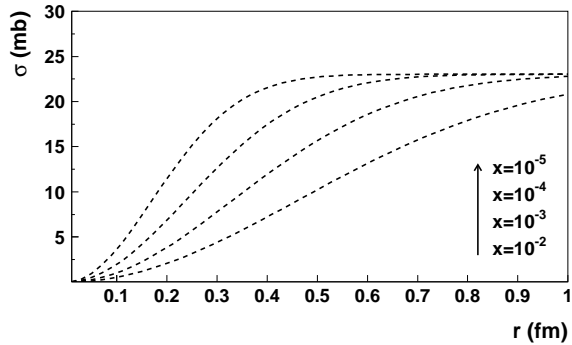,width=9cm}
           }
\vspace*{-1.0cm}
\caption{\it The dipole cross section (\ref{eq:7}).
\label{fig:dipcs}}
\end{figure}

Particularly important are those \cite{GLR} which do not lead to the violation of the Froissart unitary bound for the total $\gamma^*p$ cross section: $\sigma^{tot}\le c \ln^2 W^2$.
Applying the $q\bar{q}$ dipole picture to $\sigma^{tot}$, the following relation holds in the small-$x$ limit \cite{NN}
\be
\label{eq:6}
\sigma^{tot}\,=\,
\int d^2 r\, dz\,
|\Psi^\gamma(r,z,Q^2)|^2\ \hat\sigma(x,r),
\ee
with the same dipole cross $\hat\sigma(x,r)$ as in (\ref{eq:5}). In order to fulfil
the Froissart bound, the following phenomenological form of
the dipole cross section was proposed in \cite{GBW}
\be
\label{eq:7}
\hat{\sigma}(x,r)\,=\,\sigma_0\, \{1-\exp(-r^2 Q^2_s(x))\}\,,
\ee
where $Q_s(x)=Q_0\,x^{-\lambda}$ is a saturation scale  which parameters
(together with $\sigma_0$) were found from a fit to all small-$x$ data on
$\sigma^{tot}\sim F_2/Q^2$. Having obtained the dipole cross section from the analysis of inclusive data, it can be used to predict diffractive cross sections in DIS. This strategy
was sucessfully applied in \cite{GBW2}.

Formula (\ref{eq:7}) captures essential features of parton saturation \cite{GLR,SATREV}.
For $r\gg 1/Q_s(x)$ the dipole cross section saturates to a constant value $\sigma_0$,
which may be regarded as a unitarity bound leading to the behaviour respecting the Froissart condition: $\sigma^{tot}\sim \ln W^2$. For $x\to 0$
the dipole cross section saturates for smaller dipoles, thus with increasing  energy
the proton blacken for the dipole probe of fixed transverse size. An important aspect of the
form (\ref{eq:7}), in which $r$ and $x$ are combined
into one dimensionless variable $r Q_s(x)$, is  
geometric scaling, new scaling in  inclusive DIS  at small $x$ \cite{GSCAL}.
Qualitatively, the behaviour (\ref{eq:7}) can be found from an effective theory of dense parton systems with saturation  -- the Color Glass Condensate, see \cite{SATREV} and reference therein.

The DIS diffraction is an ideal process to study parton saturation  since it is
especially sensitive to the large dipole contribution, $r>1/Q_s(x)$.
Unlike inclusive DIS, the region below is suppressed by an additional power of $1/Q^2$.
The dipole cross section with saturation (\ref{eq:7}) leads in a natural way
to the constant ratio  (up to logarithms) \cite{GBW}
\be
\label{eq:8}
\frac{\sigma^{diff}}{\sigma^{tot}} \sim \frac{1}{\ln(Q^2/Q^2_s(x))}\,.
\ee

\begin{figure}[t]
  \vspace*{-1.0cm}
     \centerline{
         \epsfig{figure=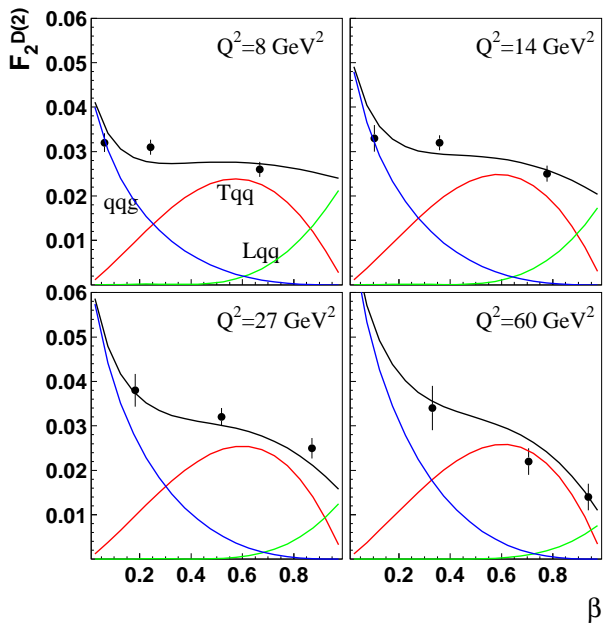,width=9cm}
           }
\vspace*{-1.0cm}
\caption{\it The diffractive structue function as a function of $\beta$
at fixed $\xp=0.003$. Three components of the diffractive system are shown.
\label{fig:ddzeus}}
\end{figure}

In the analysis \cite{GBW2} of DIS diffraction, the dipole cross section (\ref{eq:7})
was used for the description of the interaction of the diffractive system and 
the proton. The simplest system, which dominates for diffractive masses $M^2\sim Q^2$, is formed by the $q\bar{q}$ pair. However, 
for large diffractive masses, $M^2\gg Q^2$, the $q\bar{q}g$ component is more important.
In Fig.~\ref{fig:ddzeus} we show the result of the comparison of the saturation model predictions with the ZEUS data \cite{ZEUS99}, indicating three components of the diffractive system:
the $q\bar{q}$ state from transverse and  longitudinal polarized virtual photon, and  $q\bar{q}g$ component. A recent  analysis of diffractive data using the same idea but  different prescription for the dipole cross section is given in \cite{JEFF}.

The high energy formula (\ref{eq:5}) contains all powers of $1/Q^2$
(twists). Extracting the leading twist contribution from both the $q\bar{q}$ and $q\bar{q}g$ components,
the quark and gluon
DPD can be directly computed  in the saturation model \cite{GBW3}.
An exciting aspect of this calculation is the Regge factorization of the DPD,
\be
\label{eq:9}
\xp\, q^D(\xp,\beta)\,=\, Q^2_s(\xp)\,\bar{q}(\beta)\,\sim\, \xp^{-0.3}\,,
\ee
due to the form (\ref{eq:7}) with the combined variable $r\, Q_s$. The
dependence: $F^D_2\sim \xp^{1-2\alpha_\funp}$ with $\alpha_\funp\approx 1.15$, resulting from (\ref{eq:9}),
is in remarkable agreement  with the data \cite{H197,ZEUS99,ZEUS04}. Thus the Regge factorization and the dependence on energy of the diffractive DIS data are naturally explained
in the parton saturation approach. This fact emphasize importance of unitarity in the QCD description of DIS diffraction. 

\section{Soft color interactions}

\begin{figure}[t]
     \centerline{
         \epsfig{figure=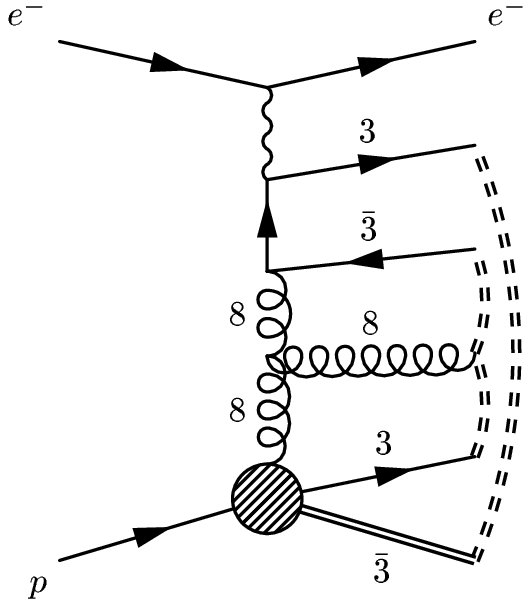,width=4cm}
         \hskip 2cm
         \epsfig{figure=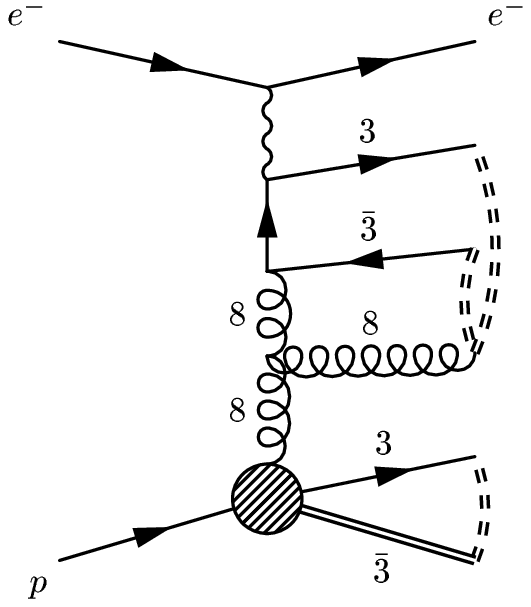,width=4cm}
           }
\caption{\it Typical string configuration in the Lund string model (left) and
confi\-gu\-ra\-tion  after the color rearrangement (right).
\label{fig:sci1}}
\end{figure}

The rapidity gaps in the diffractive interactions are explained in the discussed models by
the color singlet, vacuum exchange -- the pomeron, being a complicated gluon exchange interaction. The basic assumption in the soft color interaction model \cite{SCI} is that
the underlying hard interaction of a diffractive event is the same as in a typical DIS event.
Thus, the flattnes of the ratio $\sigma^{diff}/\sigma^{tot}$ in both $x$ and $Q^2$ is a natural
consequence of this model. The color singlet exchange responsible for the rapidity gap is the result of
soft interactions which rearrange color of the final state partons without affecting
their momenta, see Fig.~\ref{fig:sci1}. This leads to a region in phase space
without string in the Lund model of hadronization, which leads to the rapidity gap.
Such a reshuffling in color space was implemented in the Monte Carlo event generators, providing good description of the diffractive data in DIS.

\begin{figure}[h]
  \vspace*{-1.0cm}
     \centerline{
         \epsfig{figure=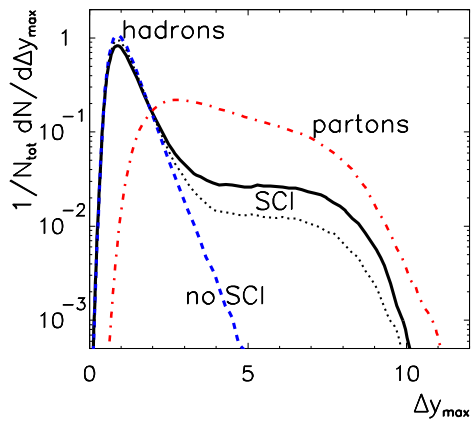,width=7cm}
           }
\vspace*{-1.0cm}
\caption{\it Distribution of events with maximal rapidity gap $\Delta y_{max}$
             in DIS events (reproduced from \cite{SCI}).
\label{fig:sci2}}
\end{figure}

In Fig.~\ref{fig:sci2} the distribution of
events with maximal rapidity gap $\Delta y_{max}$ is shown, using Monte Carlo models with
and without the soft color interactions (SCI). As we see, without  SCI large rapidity gaps
are exponentially suppressed. This is not the case for SCI.
In summary, in the presented approach the rapidity gap formation is  a final state soft effect
which is not connected to the hard scattering process.

\section{Summary}

The unexpectedly large fraction of diffractive DIS events observed at HERA renewed an interest
in diffractive phenomena in high energy scattering, now in the context of perturbative QCD.
We presented three approaches to the generation of rapidity gaps which are not
exponentially suppressed. In the first one, somewhat conventional pomeron mechanism, known from
the Regge approach to high energy scattering, was supplemented by hard structure
which emerges in the experimentally observed diffractive events. In the second approach,
DIS diffraction is strongly related to the necessity to take into account unitarization effects
in the QCD description of color singlet  gluonic exchanges. In the third approach, the difference
between the normal and diffractive DIS events lies in the final state soft interactions which
are decoupled  from the hard part of the final parton state.
All these description could be tested for more exclusive diffractive processes, e.g. in vector meson or large-$p_T$ jet production in DIS  and in hadron-hadron collisions.
For more details on hard diffraction,
we refer to the excellent reviews \cite{REVWU,HEB}.

\bigskip\bigskip
\centerline{ACKNOWLEDGEMENTS}
A partial support of the Polish State Committee for Scientific Research,  grant no. \\
1 P03B 028 28, is acknowledged.

\end{document}